# Monolayer 2D semiconducting tellurides for high-mobility electronics

Huta R. Banjade, Jinbo Pan, Qimin Yan*

Department of Physics, Temple University, Philadelphia 19122, USA

*Email: qiminyan@temple.edu

**Abstract:**

Discovery and design of two-dimensional (2D) materials with suitable band gaps and high carrier mobility is of vital importance for photonics, optoelectronics, and high-speed electronics. In this work, based on first principles calculations using density functional theory (DFT) with PBE and HSE functionals, we introduce a family of monolayer isostructural semiconducting tellurides $M_2N_2Te_8$, with M = {Ti, Zr, Hf} and N= {Si, Ge}. These compounds have been identified to possess direct band gaps from 1.0 eV to 1.31 eV, which are well suited for photonics and optoelectronics applications. Additionally, anisotropic in-plane transport behavior is observed and small electron and hole (0.11 - 0.15 $m_e$) effective masses are identified along the dominant transport direction. Ultra-high carrier mobility is predicted for this family of 2D compounds which host great promise for potential applications in high-speed electronic devices. Detailed analysis of electronic structures reveals the origins of the promising properties of this unique class of 2D telluride materials.

## Introduction

Since a decade ago, the successful fabrication of graphene[1,2], which is the first synthesized two-dimensional (2D) material, has initiated an exciting research field of 2D materials and functional devices. Great efforts have been focused on the discovery and synthesis of novel 2D materials, such as transition metal dichalcogenides (TMDCs)[3,4] and black phosphorus (BP)[5,6]. Successful fabrication of field effect transistors, photodetectors and light emitting diodes using these 2D materials have demonstrated their potential for usage in electronics. These novel 2D materials possesses remarkable properties which are absent in their bulk counterparts and have great application potential in optoelectronics, nanoelectronics and ultrathin flexible devices[7]. In addition, the presence of Dirac-cone structures and singularities in the electronic spectra of many newly predicted 2D materials make them promising for both fundamental research and potential electronic applications[8,9].

Field effect transistor (FET) is the most successful device concept and a vital component of present semiconductor electronics. Carrier mobility of the channel material and its bandgap play an important role in the performance of FETs[10] such as on-off ratio and operation speed, and the identification of novel channel materials in FET technology is a major task in semiconductor electronics. Graphene exhibits extremely high carrier mobility (10,000 – 15,000 $cm^2V^{-1}s^{-1}$ )[11], which has ignited great excitement in the device community as a possible channel material in FETs. However, the gapless electronic structure and related low on/off ratio of graphene transistors hinder their application for logic operation[12]. Huge efforts have been applied to open a sizable band gap in graphene. Unfortunately, the creation of a band gap has been always accompanied with a dramatic decrease in mobility[11].

With the successful fabrication of high-performance $MoS_2$ FETs[13], the search of novel 2D materials for electronics gained more momentum. FETs made from few layer TMDC's such as $MoS_2$ and $WSe_2$ exhibit high on/off current ratios and excellent current saturation characteristics, but the carrier mobility (< 200 $cm^2V^{-1}s^{-1}$)[14,15,16] is much smaller than that of graphene. Recent studies show that another 2D semiconductor, single layer black phosphorus (phosphorene) exhibits a high carrier mobility (~1,000 $cm^2V^{-1}s^{1}$)[6,17], but is still orders of magnitude less in comparison to graphene. Hence, the search for stable and semiconducting 2D materials as graphene alternatives that can outperform the silicon-based devices is worthwhile and will be a great boon for electronic device community.

In this work, we propose a family of $M_2N_2X_8$ type 2D monolayer semiconductors with suitable band gap, low carrier effective mass and potentially high carrier mobility along the dominant propagation direction. We perform the first principles calculations on a family of 2D ternary group-IV chalcogenides consisting of 36 $M_2N_2X_8$ compounds, with M = (Ti, Zr, Hf) from group IV, N = (Si, Ge) from group XIV and X = (S, Se or Te) from group XVI of the periodic table. Among those 36 compounds, we found that 6 telluride compounds with the chemical formula $M_2N_2Te_8$ are very promising due to their direct band gaps, strong in-plane anisotropy, and low carrier effective masses along the major transport direction.

Compounds with similar structure and composition including $ZrGeTe_4$, $HfGeTe_4$, and $TiGeTe_6$ have already been synthesized in their bulk form[18], and the measurement of their structural information and other intrinsic properties such as electrical conductivity and magnetic susceptibility of compounds have been performed[18]. Atomic structure of these semiconducting tellurides ($M_2N_2X_8$) is characterized by the presence of metal-centered bi-capped trigonal prisms as a basic structural unit, being a very common feature in ternary group-V chalcogenides, are coordinated by a "metal – metal" bond between atom M (transition metal) and N (group XIV), as shown by the relaxed structure of $Ti_2Si_2Te_8$ in Figure 1. Owing to this unique bonding structure, these 2D semiconducting tellurides exhibit many promising electronic properties which are of vital importance in applications such as optoelectronics[19], nano-electronics[7] and ultra-thin flexible devices[19,20].

## Computational details

The geometry optimization and electronic structure calculations are performed by using density functional theory (DFT) as implemented in Vienna ab initio simulation package (VASP)[21]. The projected-augmented-wave (PAW) pseudopotentials [22,23] are used to describe the valance electron and core interactions. Generalized gradient approximation (GGA)[24,25] and the Perdew-Burke-Ernzerhof (PBE)[26] exchange correlation functional are used to describe the electron exchange correlation. DFT calculations are performed at the PBE level to identify the trends in band structures and extract effective masses, through which we identify six telluride compounds with high carrier mobilities based on their band structures and effective masses. It is well known that the local/semi-local functional based DFT method severely underestimates the band gaps of semiconducting or insulating solids[27]. To partially address the band gap problem, all the electronic structure calculations on the six telluride compounds are performed by using the Heyd-Scuseria-Ernzerhof hybrid functional (HSE)[28,29]. A $9\times 3\times 1$ Γ-centered k-point mesh and a planewave energy cut-off of 400 eV are used which provide well converged results. All atoms are relaxed until the final force exerted on each atom is less than 0.01 eV/Å and the change in total energy between the two steps is less than $10^{-4}$ eV.

Fig. 1(b) and 1(c) illustrate the typical electronic band structure and atomic projected density of states (PDOS) of $Ti_2Si_2Te_8$ as an outstanding example of the ternary 2D telluride family (see band structures for all other compounds in the Supplementary Materials). Detailed study of band structure shows that, conduction band minima (CBM) and valance band maxima (VBM) lie at the center of Brillouin zone, i.e. Γ point, indicating a direct transition between CBM and VBM is plausible. Our hybrid functional calculations show that there are sizable direct band gaps in these 2D compounds, ranging from 1.00 eV in $Ti_2Ge_2Te_8$ to 1.31 eV in $Hf_2Si_2Te_8$. Predicted values of band gaps, along with relaxed lattice constants, are presented in Table 1. Large dispersion in both conduction and valences bands close to Γ indicates the lower carrier effective mass along the dominant transport direction. Because of the variation in dispersion in band along the dominant transport direction, anisotropic effect can be observed in the band structure and hence the anisotropic transport behavior of charge carriers is expected.

Electron and hole effective masses for these 2D telluride compounds are estimated by using the parabolic fitting of the energy bands around CBM and VBM near Γ point in *k* space. Effective masses of electrons and holes span a wide range from $0.11m_0$ to $1.26m_0$ along the dominant transport direction, where $m_0$ is electron rest mass. Predicted values of effective masses for electrons and holes along the Γ-X and Γ-Y directions are presented in Table 2. Variation in the calculated effective masses along the two in-plane directions indicate the strong anisotropic electric transport in these materials. This anisotropic transport behavior is similar to that of black phosphorene[6]. The PDOS in these compounds shows that the valance band maxima is mostly dominated by Te *p*-states, while the conduction band minima is mainly constructed by transition metal (i.e. M) *d*-states, with the presence of strong hybridization between atom M and N in some compounds. For instance, Figure 1(c) shows the PDOS for $Ti_2Si_2Te_8$, in which the valance band maxima are dominated by Te *p*-states while the conduction band minima are dominated by Ti *d*-states. Consistent with this observation, the band decomposed partial charge density (Fig. 2) for the electronic state at the VBM and CBM in $Ti_2Si_2Te_8$ is localized at Te and Ti atoms respectively. Detailed analysis of band decomposed partial charge densities for all other compounds are presented in the supplementary material.

Electronic properties of mono to several layer 2D semiconductors are mainly governed by carrier mobilities. We estimate theoretically the carrier mobilities for the family of 2D tellurides along the two in-plane transport directions (Γ-X and Γ-Y) based on the theoretical approach proposed by J. Bardeen *et al.*[30] with a phonon-limited scattering model in which carrier mobility is primarily limited by the scattering due to phonons[30]. Due to the inverse relationship between mobility and effective mass, a small effective mass is obviously one of the preliminary requirements for high carrier mobility. In addition to carrier effective mass, other important factors affecting the mobilities include the deformation potentials and the elastic modulus along the propagation direction of longitudinal acoustic waves[6].

In this work, the carrier mobility of 2D materials is estimated by a simplified relation[6]: $\mu_{x(y)} = \frac{e\hbar^3 \; C_{x(y)2D}}{k_B T m_e^* m_d (E_1^i)^2}$, where, $m_\mathrm{d} = \sqrt{(m_x^{i*} m_y^{i*})}$ is the average effective mass, *i* represents electron for the conduction band or hole for the valance band (subscript *x* and *y* indicate the Γ-X and Γ-Y transport directions), $m_{e*}$ is the carrier effective mass, T is the temperature (room temperature T = 300K is used), $E_{1i}$ stands for the deformation potential along the transport direction, and $C_{x(y)2D}$ is the 2D elastic modulus along the transport direction. Deformation potential for both electrons and holes along *x* and *y* direction are obtained by the linear fitting of band energy at the CBM and VBM with respect to strains along Γ-X and Γ-Y. 2D elastic modulus are calculated using the relation, $(E-E_0)/S_0 = C_{x(y)2D} (\Delta l/l_0)^2/2$, where $E_0$ is the total energy and $S_0$ is the lattice area at equilibrium for the 2D system, $l_0$ is the equilibrium lattice constant along the transport direction and $\Delta l$ is its change due to strain . The calculation details of deformation potential, 2D elastic modulus and fitted curves for change in band edge and total energy for all compounds are presented in Figure S1 of the supplementary information.

Computed values of effective masses, deformation potentials, 2D elastic modulus and carrier mobilities for all the 2D telluride compounds are presented in Table 2. Carrier mobilities in these compounds range from hundred to almost tens of thousands of $cm^2V^{-1}s^{-1}$, which endow this family of 2D compounds a great potential for electronic and optoelectronic applications. These carrier mobilities exhibit high in-plane directional anisotropy, with electrons being more mobile in general. Even though the electron effective mass (0.33 $m_0$ in $Ti_2Si_2Te_8$ and 0.19 $m_0$ in $Zr_2Si_2Te_8$) along Γ-X direction is slightly larger compared with other known 2D compounds for electronic applications, the computed electron mobility along *x* direction is extremely large in $Ti_2Si_2Te_8$ and $Zr_2Si_2Te_8$ due to rather small absolute deformation potentials for conduction band along *x* direction. The computed results show that holes are more mobile along Γ-Y (except in $Hf_2Ge_2Te_8$).

As carries mobilities are highly correlated with the deformation potentials, the understanding on the anisotropy in deformation potentials is essential. Since the computed electron mobility in these compounds is higher along Γ-X direction, here we focused ourselves in understanding the difference in deformation potential related to strain along *x* direction via the VBM and CBM wavefunctions and bonding analysis between neighboring atoms utilizing the projected Crystal orbital Hamiltonian population (pCOHP) as incorporated in the LOBSTER package[31]. pCOHP is a powerful physical quantity to understand the details of bonding, nonbonding and antibonding interactions between pair of atoms and their atomic orbitals in a compound[32]. Here we present the -pCOHP in which a positive value corresponds to the bonding state and negative value to the antibonding state.

As shown in Fig. 3, taking $Ti_2Si_2Te_8$ as an example, there exists a strong bonding interaction between Ti/Si atoms and neighboring Te atoms along *x* direction at the VBM, while a mixture of bonding and antibonding interaction is observed at the CBM. The VBM wavefunction in Fig. 2(a) shows a strong overlap along *x* direction, indicating that a small structural deformation along *x* direction may have a remarkable effect on this electronic state and hence cause a significant change in its energy, resulting in a large deformation potential. The situation is rather different for the CBM wavefunction along *x* direction which exhibits much weaker orbital overlap (Fig. 2(b)). As observed in Fig. 3(b), the mixture of bonding and antibonding interaction between neighboring atoms along *x* direction results in a dramatic decrease in overall strength of the orbital interaction. Due to this effect, structural deformation has a much less effect on the CBM wavefunction along *x* direction, which results in the smaller values of deformation potential. Along the *y* direction, the VBM wavefunctions (as shown in Fig. 2(a)) are more localized than the CBM wavefunctions (Fig.

2(b)), resulting in smaller values of deformation potential at the VBM than those at the CBM along *y* direction. As the nature of anisotropy in the observed deformation potentials is similar in these compounds, we conclude that this anisotropic transport behavior is dominated by the bonding and antibonding interaction between the neighboring atomic orbitals and their different response under strain. Mobility along a transport direction is inversely proportional to the square of the deformation potential along that direction. Therefore, the mobility along a transport direction is strongly controlled (but not completely determined) by the deformation potential in that direction. The observed higher electron mobilities along *x* direction in these telluride compounds are mostly due to the lower values of deformation potential at the CBM along *x* direction.

In summary, we predict a family of 2D ternary semiconducting tellurides with the chemical composition $M_2N_2Te_8$ by using first-principles computations based on density functional theory. Computed band gaps and carrier mobilities of this compound set exhibit great potentials for electronic and optoelectronic applications. All of these compounds possess direct band gaps at the center of Brillouin zone (i.e. Γ point) with extremely high electron mobilities higher than that of single layer BP, $MoS_2$, and other 2D $MX_2$ semiconductors. Out of these six compounds, we present two benchmark systems $Ti_2Si_2Te_8$ with a band gap at 1.03 eV and electron mobility at $9.92 \times 10^3$ $cm^2V^{-1}s^{-1}$ and $Zr_2Si_2Te_8$ with a band gap at 1.24 eV and electron mobility at $5.58 \times 10^3$ $cm^2V^{-1}s^{-1}$, which are promising for logical devices that require high mobility and optimal band gap. Similar to single layer BP, the observed anisotropy in mobility in these compounds is mainly due to the anisotropy in deformation potential, which is correlated with the bonding-antibonding interactions between neighboring atomic orbitals and subsequently different response of the CBM and VBM under strain. The computational identification of these 2D telluride compounds provides a novel material platform for experiments in the search for functional materials that enable future 2D electronic devices.

**Acknowledgement**

This work was supported as part of the Center for Complex Materials from First Principles (CCM), an Energy Frontier Research Center funded by the U.S. Department of Energy (DOE), Office of Science, Basic Energy Sciences (BES), under Award DE-SC0012575. H.B. and Q.Y. acknowledge the support by the U.S. Department of Energy, Office of Science, under award number DE-SC0020310. It benefitted from the supercomputing resources of the National Energy Research

Scientific Computing Center (NERSC), a U.S. Department of Energy Office of Science User Facility operated under Contract No. DE-AC02-05CH11231, and Temple University's HPC resources supported in part by the National Science Foundation through major research instrumentation grant number 1625061 and by the US Army Research Laboratory under contract number W911NF-16-2-0189.

**Table 1:** Relaxed lattice constants (*a* and *b*) and computed band gaps estimated by using the HSE06 functional.

| **Compounds** | **Lattice constant** (Å) | | **Band gap (eV)** |
|---|---|---|---|
| | ***a*** | ***b*** | |
| **$Hf_2Ge_2Te_8$** | 3.99 | 10.99 | 1.29 |
| **$Hf_2Si_2Te_8$** | 3.96 | 10.85 | 1.31 |
| **$Ti_2Ge_2Te_8$** | 3.90 | 10.76 | 0.99 |
| **$Ti_2Si_2Te_8$** | 3.87 | 10.61 | 1.03 |
| **$Zr_2Ge_2Te_8$** | 4.01 | 11.04 | 1.23 |
| **$Zr_2Si_2Te_8$** | 3.97 | 10.90 | 1.24 |

**Table 2:** Effective mass $m_x^*$ ($m_y^*$), deformation potential $E_{1x}$ ($E_{1y}$), 2D Elastic modulus $C_{x2D}$ ($C_{y2D}$) and the carrier mobility $\mu_x$ ($\mu_y$) along the dominant transport direction Γ-X and Γ-Y for electron (e) and hole (h) carriers.

| Compound | | $m_x^*$ ($m_0$) | $m_y^*$ ($m_0$) | $E_{1x}$ (eV) | $E_{1y}$ (eV) | $C_{x2D}$ ($Jm^{-2}$) | $C_{y2D}$ ($Jm^{-2}$) | $\mu_x$ ($10^3$ $cm^2V^{-1}s^{-1}$) | $\mu_y$ ($10^3$ $cm^2V^{-1}s^{-1}$) |
|---|---|---|---|---|---|---|---|---|---|
| $Hf_2Ge_2Te_8$ | e | 0.28 | 0.35 | 3.55 | 3.58 | 87.25 | 45.75 | 1.70 | 0.70 |
| | h | 0.11 | 1.13 | 9.00 | 2.27 | 87.25 | 45.75 | 0.60 | 0.47 |
| $Hf_2Si_2Te_8$ | e | 0.25 | 0.43 | 2.58 | 2.29 | 77.21 | 53.41 | 3.10 | 1.54 |
| | h | 0.12 | 1.01 | 9.85 | 1.99 | 77.21 | 53.41 | 0.40 | 0.82 |
| $Zr_2Ge_2Te_8$ | e | 0.30 | 0.32 | 3.62 | 3.03 | 77.49 | 39.17 | 1.36 | 0.92 |
| | h | 0.12 | 1.21 | 9.37 | 2.02 | 77.49 | 39.17 | 0.41 | 0.45 |
| $Zr_2Si_2Te_8$ | e | 0.29 | 0.44 | 1.68 | 2.01 | 77.47 | 50.46 | 5.58 | 1.68 |
| | h | 0.15 | 1.12 | 9.52 | 1.90 | 77.47 | 50.46 | 0.30 | 0.65 |
| $Ti_2Si_2Te_8$ | e | 0.33 | 0.29 | 1.28 | 3.46 | 77.40 | 62.50 | 9.92 | 1.24 |
| | h | 0.14 | 1.07 | 9.79 | 2.55 | 77.40 | 62.50 | 0.32 | 0.49 |
| $Ti_2Ge_2Te_8$ | e | 0.38 | 0.22 | 2.10 | 4.30 | 67.83 | 51.69 | 2.98 | 0.94 |
| | h | 0.13 | 1.26 | 9.848 | 2.47 | 67.83 | 51.69 | 0.29 | 0.35 |

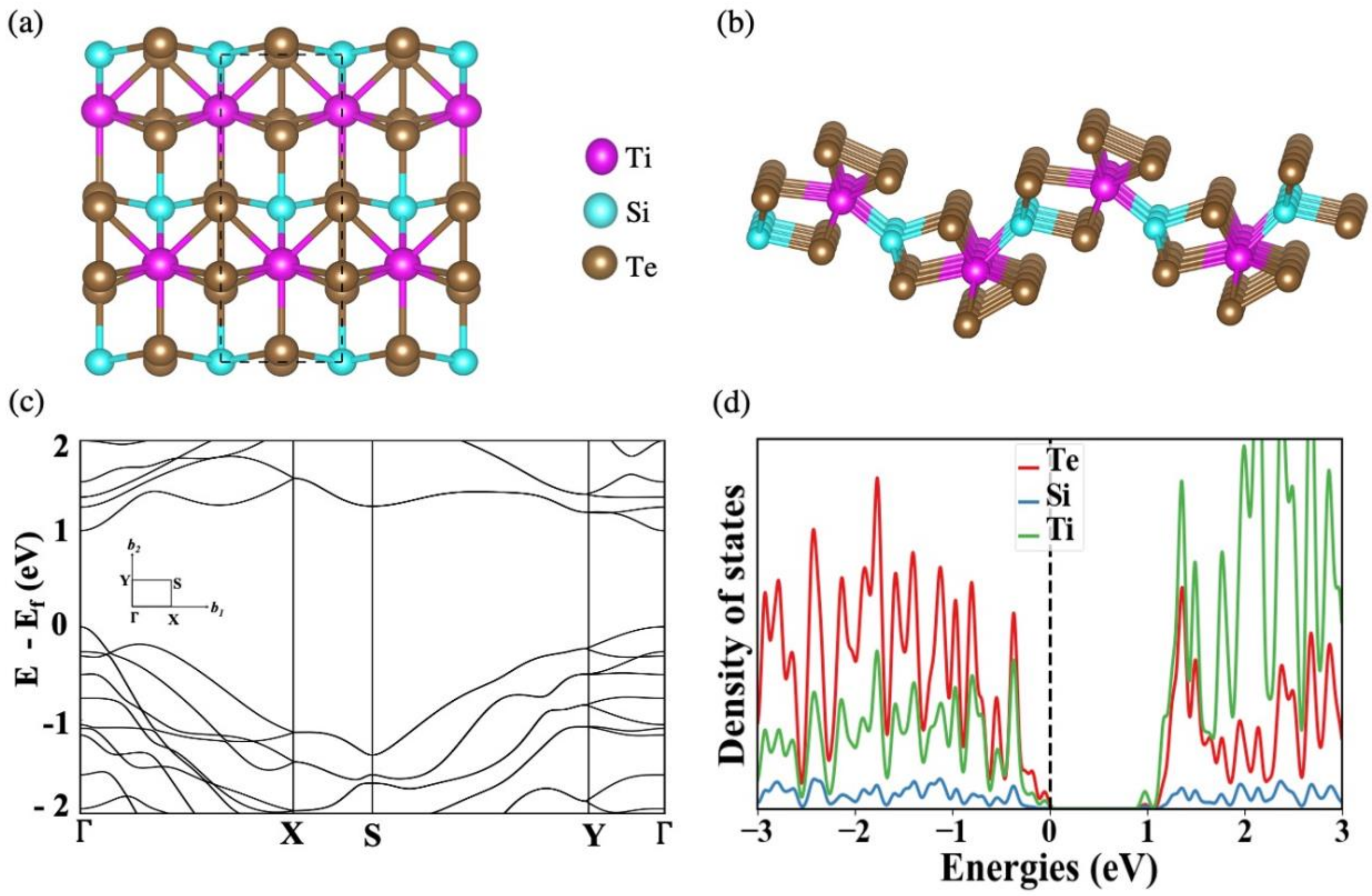

Figure 1: (a) Top and (b) side view of the relaxed structure of $Ti_2Si_2Te_8$. The magenta, cyan and brown balls represent Ti, Si, and Te atoms, respectively. (c) Band structure of $Ti_2Si_2Te_8$ calculated on the level of HSE06 hybrid functional. High-symmetry k points in the reciprocal space are presented in the figure. (d) Projected density of states of $Ti_2Si_2Te_8$, showing that the dominant states at the CBM and VBM are contributed by Ti *d*-states and Te *p*-states, respectively.

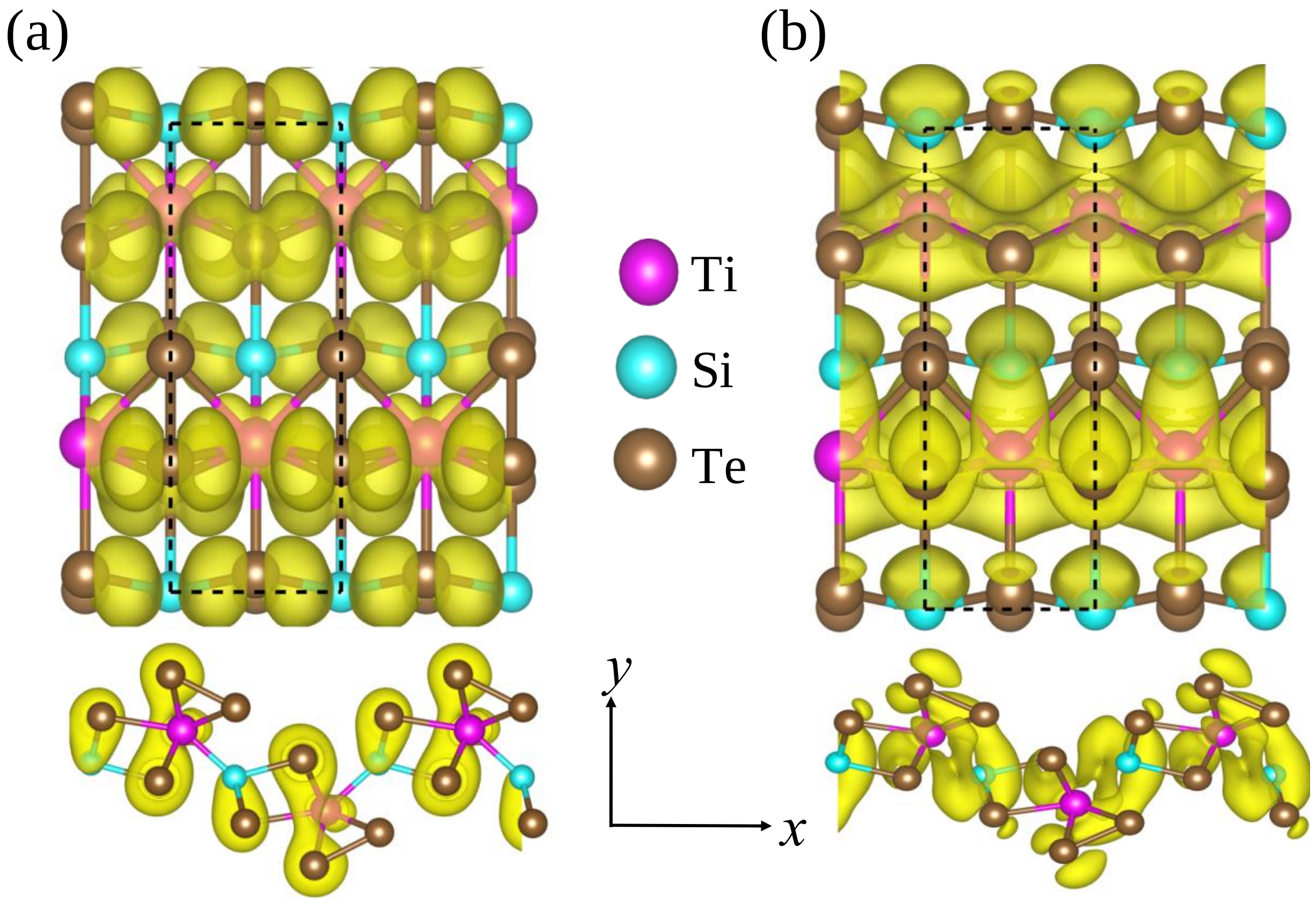

Figure 2: Top and side view of band decomposed partial charge density of $Ti_2Si_2Te_8$ at (a) the VBM and (b) the CBM at Γ point in *x*-*y* plane at the 0.0064 $e\text{Å}^{-3}$ isosurface level. The magenta, cyan and brown balls represent Ti, Si, and Te atoms, respectively

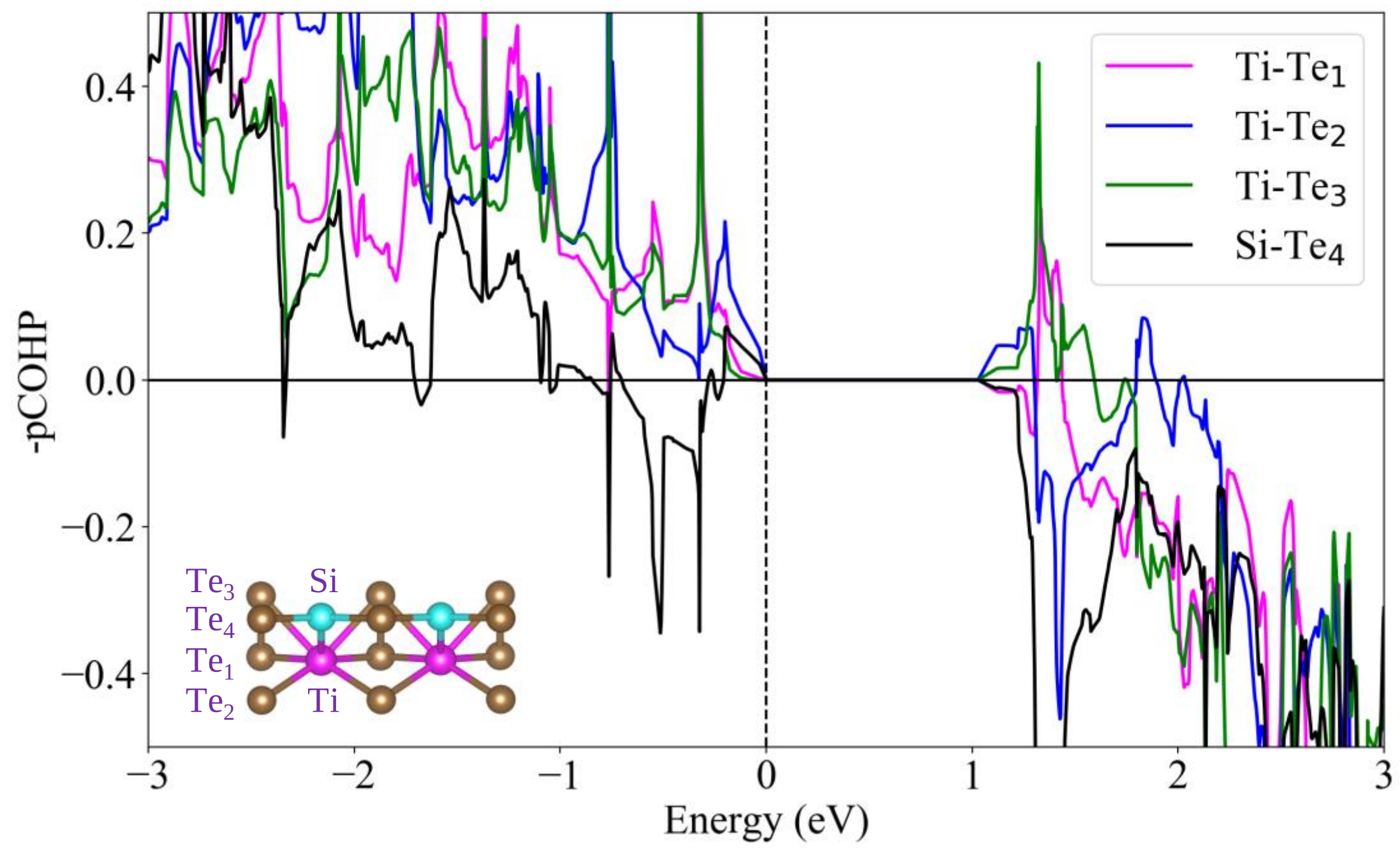

Figure 3: Projected crystal orbital Hamiltonian population (-pCOHP) between neighboring atoms of $Ti_2Si_2Te_8$, with the corresponding structure showing the neighboring atoms of which pCOHP is obtained from.